\documentclass[twocolumn,
showpacs,
amsmath,
amssymb,
amsfonts,
a4paper,
aps,
citeautoscript,
footinbib,
pra,
superscriptaddress]{revtex4}
\usepackage{times}
\usepackage{graphicx}
\usepackage[latin1]{inputenc}

\graphicspath{{./Figs/}}

\begin{document}
\author{Guillaume Roux}
\email{guillaume.roux@u-psud.fr}
\affiliation{Univ Paris-Sud, Laboratoire de Physique Th\'eorique et Mod\`eles Statistiques, UMR8626, Orsay F-91405, France;
CNRS, Orsay, F-91405, France.}

\date{\today} 

\title{Reply to ``Comment on `Quenches in quantum many-body systems: One-dimensional Bose-Hubbard model reexamined' ''}


\begin{abstract}
In his Comment [see preceding Comment, Phys. Rev. A 82, 037601 (2010)] on the paper by Roux [Phys. Rev. A 79, 021608(R) (2009)], Rigol argued that the energy distribution after a quench is not related to standard statistical ensembles and cannot explain thermalization. The latter is proposed to stem from what he calls the eigenstate thermalization hypothesis and which boils down to the fact that simple observables are expected to be smooth functions of the energy. In this Reply, we show that there is no contradiction or confusion between the observations and discussions of Roux and the expected thermalization scenario discussed by Rigol. In addition, we emphasize a few other important aspects, in particular the definition of temperature and the equivalence of ensemble, which are much more difficult to show numerically even though we believe they are essential to the discussion of thermalization. These remarks could be of interest to people interested in the interpretation of the data obtained on finite-size systems.
\end{abstract}

\maketitle 

As an introduction, we briefly summarize our point of view regarding Rigol's Comment and then discuss in more details the arguments supporting it. We do agree with Rigol that the fine structure of the energy distribution may not, in principle, affect the thermalization scenario after a quantum quench. We also agree that the ETH would participate in explaining why simple observables can look thermalized after a unitary evolution from an initial state. However, this is expected to be true only in the thermodynamical limit, and provided the model is well-behaved (in this respect). We show below that there is no contradiction nor confusion with the statements of Ref.~\onlinecite{Roux2009} which are correct on finite systems and, we believe, actually relevant to interpret the BHM numerical data at stake. We lastly point out some difficulties with the interpretation of the data of Rigol's Comment, basically that there is no use of the microcanonical entropy to define the temperature because ensemble equivalence is not reached. In this respect, Ref.~\onlinecite{Roux2009} provides evidence that the Shannon entropy of the time-averaged density-matrix, on finite systems, depends on the initial state (something that may disappear in the thermodynamic but is much more difficult to prove numerically than checking the ETH on simple observables).

As a preamble, we recall a general result on statistical ensembles and their equivalence, something important in the context of quantum quenches and relevant to the present discussion. On finite systems, the microcanonical and canonical ensembles are not equivalent and will lead to different predictions. The two ensembles lead to identical predictions for the entropy v.s. mean-energy relation (useful for thermodynamics) only in the thermodynamical limit and under some rather general assumptions  (scaling of the density of states with the number of particles, behavior of the energy fluctuations and mean-energy of the system) allowing for a saddle-point approximation of the energy distribution. This is called the ensemble equivalence (EE). From these arguments, which can be found in usual statistical mechanics textbooks~\cite{StatMech}, it sounds sensible to think that any well-behaved diagonal ensemble (coming from a ``generic'' initial state, including the possibility of a quench process) will be equivalent to both the microcanonical and canonical ensembles, leading to the same thermodynamics. Statistical mechanics is mostly focused on the energy distribution as it tells, independently of the behavior of observables, which states are relevant to the physics, or in a more semi-classical approach, which parts of the phase space contribute and how.

In addition to the energy distribution, the behavior of the observables with energy is also an important issue when one wants to average observables over statistical ensembles. The fact that observables should behave smoothly with the energy is sometimes called the semi-classical approximation (SA) and appears in standard statistical physics textbooks~\cite{StatMech} on phenomenological grounds, when one needs to go from a sum over micro-states to an integral over energy. What is called ETH by Rigol is based, in our opinion, on these two phenomenological ideas that seem to help explain thermalization in a closed quantum system \emph{only in the thermodynamical limit} and that are certainly correct for most ``generic'' systems. The question is rather one can provide evidences or proofs supporting these arguments. In this respect, we point out some analytical work supporting SA~\cite{Peres1984, Deutsch1991, Srednicki1994, Biroli2009} and numerical simulations on small systems~\cite{Feingold1984, Jensen1985, Rigol2008, Rigol2009}. As numerics are done on finite systems, one is not in the regime of validity of both the EE and SA and one has to try to understand the finite size effects in order to give convincing data supporting the ETH. In addition, equilibrium statistical mechanics is not meant to explain only thermodynamics, it also describes fluctuations and finite size effects that are essential for many systems, including experimental ones such as cold atoms. Hence, the features of the diagonal distribution other than the mean-energy can be physically relevant and interesting in themselves, motivating the discussion of their shape.

The motivation of Ref.~\onlinecite{Roux2009} was to discuss the behavior of the diagonal ensemble only and to leave the discussion of the ETH for a later study~\cite{Roux2009a} since comparison between time-averaged observables and equilibrium predictions were already available~\cite{Kollath2007}. It provided the bare data of the distribution to see whether they were supporting or not results of Ref.~\onlinecite{Kollath2007}, which could have been spoiled, for example, by the use of a finite time window. Other motivations were to systematically investigate the variation of the quench parameter (which tunes the mean-energy of the system), to look at finite size effects, the energy fluctuations and Shannon entropy of the diagonal ensemble. The manuscript did not intend to make general claims about a thermalization mechanism (something on which the first sentence of Rigol's Comment could shed some confusion). It is never mentioned that a Boltzmann distribution is expected in general and in the thermodynamical limit, and that this would be the explanation of thermalization in a generic closed quantum system. If the title, abstract and conclusion were not clear enough, this Reply clarifies the motivations. 

We now turn to the observation of an ``approximate'' Boltzmann law and Rigol's criticism. The first point is whether it is confusing or not to state that observing a Boltzmann-like (or exponential-like) behavior for the diagonal weights supports the observation of a ``thermalized'' regime as in Ref.~\onlinecite{Kollath2007} (notice the quotation marks, as in the manuscript). Although we agree that we would have expected that the shape does not matter, we have strong finite size effects (large weight on the targeted ground-state and mean-energy close to the ground-state~\cite{Roux2009a}) and then, the ensembles are not equivalent (they would give different observables). In this case, the shape of the finite-size distributions are crucial to understand the time-averaged observables calculated previously. Notice that the comparison in Ref.~\onlinecite{Kollath2007} was carried out using quantum Monte-Carlo, i.e. using \emph{a (grand)-canonical ensemble}. The statement of Ref.~\onlinecite{Roux2009} is that this observation on finite systems supports the previously obtained results, carried out with similar system sizes, in an approximately independent way of how behave the observables. Consequently, there is no contradiction with the ETH and Rigol's argument (or EE) is actually not relevant to explain the observed ``thermalized regime'' of the BHM on finite size systems (in our opinion). In Ref.~\onlinecite{Roux2009a}, we give another \emph{example} of a quench distribution displaying a Boltzmann-like behavior on finite size systems in the perturbative regime, based on analytical calculations. 

Rigol's Comment introduces a unique definition of the effective temperature while it was not discussed in Ref.~\onlinecite{Roux2009}. We would like to stress that the introduction of an effective temperature is actually involved and shows the limitations of Rigol numerical ``proof'' of the ETH. Firstly, for the parameters of the BHM where the distributions are Boltzmann-like, one can ask whether the temperature is well defined due to the presence of finite size effects: the answer is no~\cite{Roux2009a}, i.e. different observables or definitions of temperature yield different results. We don't know how the distributions will evolve when one works on very large systems and is yet in the perturbative regime; there is no claim on this in Ref.~\onlinecite{Roux2009}. 
Secondly, similar finite size effects occur in Rigol's data: the definition used is valid only in the thermodynamical limit. Indeed, the effective temperature is actually taken from the \emph{canonical ensemble}. Thus, EE is implicitly assumed so that the temperature would have to agree with the one obtained using $1/T = \partial S/\partial E$ using Boltzmann's formula $S = k_B\ln \Omega$ and $\Omega$ given by the microcanonical ensemble. However, we see from the fluctuations (for instance) that the EE is not reached in the data of Ref.~\onlinecite{Rigol2008} and in the Comment. The advantage of using the canonical ensemble to define an effective temperature is that it is a continuous and increasing function of the mean-energy. As discussed in Ref.~\onlinecite{Roux2009}, the statistical entropy of the diagonal ensemble on a finite system is different for two mixed states with the same mean-energy. The reason is that the statistical/Shannon entropy is much more sensitive to the details of the distribution because of the log term. In Rigol's Comment, we may guess that the two data sets with the same mean-energy will have different statistical entropies and different effective microcanonical temperatures. Another possible definition of the temperature, more likely to be useful for experiments, would be to fit the momentum distribution $n(k)$ using the canonical ensemble with $T$ as a free parameter. In Ref.~\onlinecite{Rigol2008} and in the Comment, this would certainly give another different temperature. Consequently, one cannot consider Rigol's data as corresponding to a fully thermalized system either, although the ETH arguments are certainly reasonable.
A last striking difference between the data of the Comment and the perturbative regime of the BHM is that the mean-energy are at very different places: in the bulk of the spectrum for the first set and very close to the ground-state for the second. The second regime is expected to have stronger finite-size effects~\cite{Roux2009a}.

Finally, there is another issue when discussing thermalization with putting the emphasis only on the SA as in the Comment. To give an exaggerated picture of the ETH, consider the situation where a ``simple'' observable has a totally flat behavior with the energy: any distribution in phase space would lead to thermalization according to the mere comparison of observables. Hence, the energy distribution in itself is a central object. In Fig. 3(b) of Ref.~\onlinecite{Rigol2008} and in the figure of the Comment, we see that, although the ensemble are not equivalent, the comparison between the averaged observables gives very close results because the observable varies smoothly enough. In other words, the finite size effects on checking the SA/ETH are here smaller than the one on checking the EE. Thus, the discussion on a finite system in Rigol's approach can be very observable-dependent (which is well discussed in Ref.~\onlinecite{Rigol2008} on the basis of general arguments and earlier results from the literature, but not addressed numerically/quantitatively).

In conclusion, showing the bare data of the diagonal ensemble distribution corresponding to the BHM is, in our view, important to explain the numerical findings. There is no contradiction nor confusion between the statements of Ref.~\onlinecite{Roux2009} and the ETH. Rigol's Comment, although very interesting, does not bring clarifications on the particular issues of the BHM. The Comment actually reuses numerical results and ideas that could already be found in Refs.~\onlinecite{Rigol2008} and \onlinecite{Rigol2009} (up to an insignificant change in one parameter). The current numerical investigations, though they provide remarkable results, yield fully conclusive answers neither for hard-core bosonic models or the BHM, and are not in contradiction with each other (in our view). For instance, the thermalization of the BHM at large but finite U on very large systems is still an open question, although the default answer is that we expect that it occurs.

\end{document}